[Digite aqui]

# Utilizando o captador de guitarra como sensor: Construindo um *kit* para abordar a Lei de Faraday em sala de aula.

Autores:

Hugo Luna e Daniel Avila

Instituto de Física, Universidade Federal do Rio de Janeiro, Cidade Universitária, Rio de Janeiro, RJ, Brasil.

**Resumo**

A abordagem do conceito de indução eletromagnética no ensino de física, pode apresentar desafios, em particular, ao que diz respeito ao distanciamento do cotidiano dos alunos. O funcionamento de um captador de guitarra elétrica pode ser um exemplo interessante para contextualizar o tema permitindo aos alunos compreenderem, a correlação entre a variação do fluxo do campo magnético e a força eletromotriz induzida.

O objetivo deste trabalho é apresentar um projeto de custo acessível utilizando o captador de uma guitarra como sensor para abordar a Lei de indução de Faraday.

**Palavras-chave:** Lei de Faraday; Indução eletromagnética; captador de guitarra; Ensino de física.

**Abstract**

The approach to the concept of electromagnetic induction in physics education can present challenges, particularly regarding its distance from students' everyday experiences. The functioning of an electric guitar pickup can serve as an interesting example to contextualize the topic, allowing students to understand the correlation between the variation in magnetic field flux and the induced electromotive force.

This study aims to present a low-cost project to explore Faraday's Law of Induction through an experimental kit composed mainly of an electric guitar pickup, an electric motor with a speed controller, a Light Dependent Resistor (LDR) sensor, and permanent magnets.

1. Introdução.

A indução eletromagnética é um conceito fundamental na física e tem diversas aplicações práticas em dispositivos e tecnologias do cotidiano, desde a geração e transmissão de energia elétrica até carregadores de celulares sem fio. No entanto, o ensino desse tema pode apresentar desafios, uma vez que os mecanismos e conceitos envolvidos nem sempre são facilmente evidentes no dia a dia dos alunos. De fato, faz parte do cotidiano usar pilhas ou baterias para acender a lâmpada de uma lanterna, por exemplo. Por outro lado, não é obvio, que o mesmo resultado pode ser obtido fazendo-se variar o fluxo do campo magnético de um imã permanente (ou eletroímã), embora hoje em dia os carregadores *wireless* sejam utilizados no carregamento de celulares. O conceito de variação de fluxo de um campo vetorial não é intuitivo e requer uma capacidade de abstração geométrica que somente figuras esquemáticas apresentadas em livros didáticos nem sempre são suficientes para a compreensão do tema.

O uso de novas tecnologias, como por exemplo o uso dos sensores dos aparelhos celulares, facilita a inclusão de diferentes abordagens pedagógicas permitindo o aluno questionar e construir seu conhecimento a partir da observação e da experimentação (VIEIRA e AGUIAR, 2016). A contextualização também é um aspecto importante no ensino de física e muitas vezes pode ser desafiador encontrar exemplos que sejam ao mesmo tempo cotidianos e interessantes (CASTRO, 2007; LIMA, 2020). Neste sentido, a utilização do sensor da guitarra que traduz vibração mecânica em sinal elétrico, oferece uma oportunidade para abordar temas de física como: indução eletromagnética, ondas etc. O fato de escolher um sensor que é parte fundamental do funcionamento da guitarra, estabelece uma contextualização para a abordagem da Lei de Faraday. Uma vez que a guitarra é um instrumento que está presente em diversos gêneros musicais e é de interesse dos jovens.

Podemos encontrar na literatura diversos trabalhos relacionados ao ensino da Lei de Faraday, dos quais destacamos os trabalhos de (LIMA, 2020) e (REIS, 2018) que possuem uma abordagem voltada especificamente para o ensino médio. LIMA (2020) desenvolveu um produto educacional em forma de uma sequência de ensino investigativa, enquanto REIS (2018) empregou uma abordagem experimental da Lei de Faraday utilizando smartphone e computador.

No entanto, é importante destacar que utilizando o captador da guitarra elétrica como sensor, encontramos poucos trabalhos voltados para o ensino da Lei e Faraday. Os trabalhos de (LEMARQUAND *et al*., 2007) e (HORTON e MOORE, 2009), por exemplo, descrevem a modelagem matemática do campo magnético e da variação do fluxo do campo magnético do captador de uma guitarra. Ambos com abordagens avançadas voltadas exclusivamente para o ensino superior. AGUILAR *et al.* (2019) usaram o captador da guitarra como sensor para analisar a vibração de uma barra em diferentes configurações, obtendo por exemplo, a velocidade de propagação da onda na barra. Em outro contexto, (SANTOS *et al.* 2013), no trabalho "*Violão e guitarra como ferramentas para o ensino de física*", propuseram atividades práticas para a caracterização de propriedades do som produzido por violões e guitarras através de uma análise harmônica. Neste caso, usando a guitarra, mas tendo foco o ensino da física ondulatória.

A proposta deste trabalho é apresentar um produto didático para ser levado à sala de aula e acessível a professores do ensino médio e superior. Para alcançar esse objetivo, propomos a utilização do captador de guitarra elétrica como sensor. O captador, pode ser obtido desmontando uma guitarra ou comprando um modelo de custo acessível. Além do captador, utilizamos imãs permanentes, um motor elétrico e controlador de velocidade de rotação do motor para fazer a variação do fluxo do campo magnéticos, 2 palitos de picolé e tubo PVC para suporte, uma placa Arduino, um sensor de luz do tipo LDR para obtenção da velocidade de rotação do motor. Esse *kit* pode ser utilizado como um apoio ao professor na elaboração de atividades que permitam aos alunos compreenderem a relação entre o aparecimento de uma força eletromotriz induzida (f.e.m.) e a variação do fluxo do campo magnético.

Na seção 2 apresentamos os elementos básicos da montagem da proposta, como estrutura e funcionamento de um captador magnético, montagem e análise dos resultados extraídos a partir do kit proposto*.* Na seção 3 fazemos a correlação da grandeza medida com o ensino da Lei de Faraday.

2. Montagem do *kit* didático.
   2.1 Funcionamento do captador magnético da guitarra.

Os captadores são conhecidos como "*o coração*" da guitarra, tal analogia decorre do fato de ele ser essencial para que a guitarra produza seu som. Na sua forma mais simples, eles são produzidos a partir de um fio de cobre esmaltado enrolado no formato de bobina ao redor de imãs permanentes, conforme ilustrado na Figura 1. Envolto em uma capa, o captador é colocado abaixo das cordas para captar suas vibrações por intermédio de sinais elétricos gerados a partir do fenômeno de indução eletromagnética (ZACZESK, 2018).

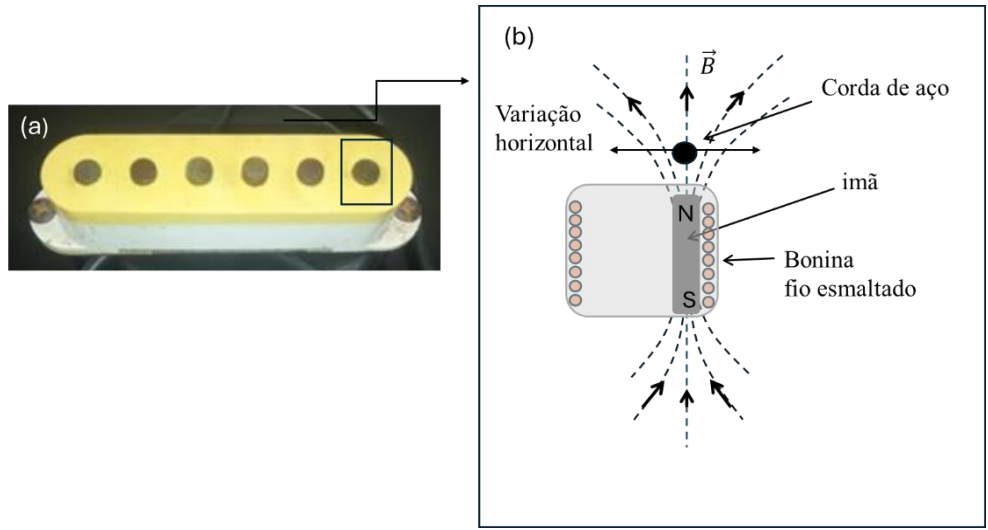

Figura 1: Esquema simplificado dos elementos de um captador magnético. (a) Foto de um captador *single coil*, (b) desenho esquemático da variação horizontal da corda da guitarra na região do imã do captador. Fonte: os autores.

As cordas da guitarra são compostas por material ferromagnético, de maneira que na presença do campo magnético dos imãs permanentes (ilustrada esquematicamente na figura 1b), passam a ser imantadas e a se comportar como imãs. Havendo vibração das cordas, o fluxo do campo magnético na localidade varia, de forma que essa vibração é transformada em sinal elétrico. Os sinais provenientes da vibração das cordas são amplificados e processados para produzir o som da guitarra de acordo com a frequência e timbre da nota musical. O funcionamento baseia-se em princípios da indução eletromagnética descritos por Faraday. Por isso, a utilização de cordas de nylon como em violões acústicos é inviável, uma vez que esse material não pode ser imantado (WERNECK 2007; LAGO 2015).

Dentre os tipos de captadores magnéticos existentes, os trabalhos de (HORTON and MOORE, 2009), e (SILVA, 2016) descrevem detalhadamente o funcionamento dos captadores destacando diferentes configurações de arranjos dos imãs permanentes. Como por exemplo, *single coil* (bobina única), *humbucker* (duas bobinas), *mini humbucker* (as bobinas são passivas e de dimensões compactas), *stacks* (que possui duas bobinas sobrepostas, sendo uma ativa e outra passiva), *dual coil* (possui duas bobinas e produz maior uniformidade do campo magnético por toda sua extensão), *tri-bucker* (associação entre três bobinas, um *humbucker* e um *single-coil*) e sistema *quad-rail* (quatro bobinas alinhadas em pares).

2.2  Montagem do *kit* didático.

Para induzir uma f.e.m. no captador da guitarra, é necessário que haja uma variação do fluxo do campo magnético a partir da vibração da corda (de material ferromagnético). Estes sinais elétricos, são tênues e necessitam de filtros e amplificação. Para contornar essa situação e evitar o uso de mais componentes elétricos, substituímos a corda por um conjunto de imãs permanentes. A variação do fluxo do campo magnético, pode ser obtida a partir da variação da posição destes imãs.

[Digite aqui]

Essa variação, pode ser feita a partir de diversas configurações geométricas, ficando a critério do professor a escolha que lhe permita a melhor aplicação da sua metodologia. Neste trabalho, nós iremos gerar a variação do fluxo do campo magnético dos imãs permanentes a partir da geometria ilustrada nas fotos da figura 2. Na figura 2a temos o captador utilizado do tipo *single coil*, escolhido por sua simplicidade e preço acessível. A variação é feita de forma periódica por conjunto de imãs permanentes que são fixados com cola, nas pontas de duas hélices (palitos de picolé) perpendiculares entre si (figura 2b). O centro das hélices está acoplado a um motor DC cuja velocidade de rotação pode variar de forma controlada até próximo ao valor máximo de 12400 rpm sem carga (~200 Hz). A f.e.m. induzida é medida por um multímetro conectado aos bornes destacados na figura 2c.

O material necessário para construção do projeto pode ser encontrado em lojas de eletrônica, de material de construção e papelaria ou comprados pela internet. Como podemos ver na figura 2, o captador está fixado no centro da face superior de uma placa MDF, posicionado logo abaixo da hélice com os imãs permanentes. O suporte do motor (tubo de PVC) também está fixado na placa MDF e a conexão entre o tubo PVC e o motor é feita por uma abraçadeira metálica.

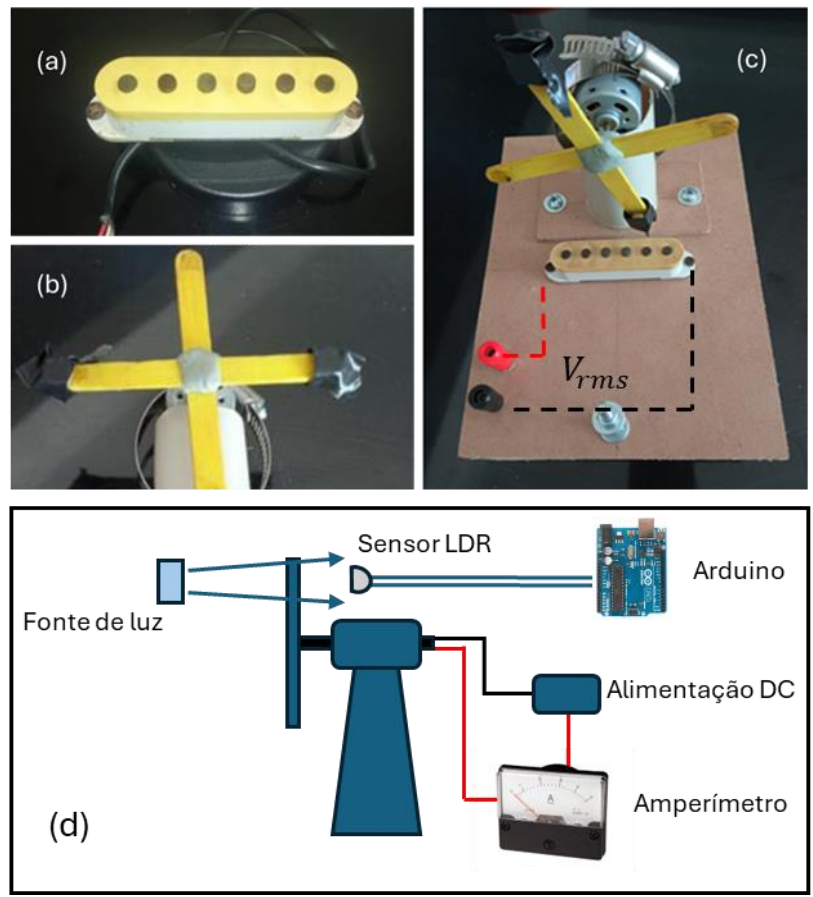

Figura 2 Foto do kit montado: (a) captador utilizado como sensor, (b) motor e hélices com imãs presos as extremidades para produzir a variação do fluxo do campo magnético, (c) todo sistema montado sobre uma plataforma com os bornes para a medição da f.e.m. induzida através de um multímetro e (d) montagem esquemática para medida da frequência de rotação do motor. Fonte: os autores.

Para alimentar o motor, usamos uma fonte genérica de notebook com chave seletora de tensão de saída (12V-24V) e corrente contínua máxima de 4,5 A. O controle de rotação foi feito com um minimódulo controlador de velocidade DC PWM com tensão de funcionamento na faixa de 5V a 35V e corrente máxima de 5A. Na tabela 1 listamos os principais componentes e os preços médios praticados no mercado Brasileiro. Os valores foram levantados utilizando sítios de busca na internet.

Tabela 1: principais componentes e preços médios encontrados em sítios de buscas na internet. Fonte: os autores.

| Componente | Preço médio em reais |
|---|---|
| Arduino UNO R3 | 60 reais |
| Arduino Nano v3 (alternativa) | 40 reais |
| Sensor LDR (5mm 5W) | 10 un. 12 reais |
| Disco imã de neodímio (D = 6mm $\times h =$ 2mm). Força de tração 600g | 10 un. 15 reais |
| Motor DC 12400 rpm | 50 reais |
| Controlador velocidade pwm (5A, 5-35V e 90W) | 30 reais |
| Multímetro digital | 30 reais |

2.3 Obtendo grandezas físicas com o *kit.*

Para relacionar a força eletromotriz induzida com a taxa de variação do fluxo do campo magnético, é necessário medir a frequência de rotação do motor. Essa medição pode ser realizada observando a saída da f.e.m. no sensor diretamente em um osciloscópio, a partir do qual é possível determinar a sua amplitude e a frequência. No entanto, o osciloscópio é um instrumento de medição de alto custo, tornando-o pouco acessível para a maioria das escolas brasileiras. Além disso, a operação deste instrumento poderia representar um nível de dificuldade adicional para os alunos.

Uma alternativa é obter a frequência de rotação de forma indireta. Essa calibração podes ser feita de diversas formas, como por exemplo: Leitura da saída digital de um sensor Hall (RIBEIRO 2024) ou vídeo análise sincronizada por uma luz estroboscópica (SILVA, 2024). Nossa opção, no entanto, foi utilizar um sensor de luz do tipo LDR (*Light Dependent Resistor*) posicionado na parte traseira do motor e hélice, enquanto este é iluminado por uma fonte de luz, como ilustrado esquematicamente na figura 2d. A cada vez que uma hélice se alinha ao LDR, uma sombra é projetada, fazendo com que a leitura da saída seja alterada. Para medir e armazenar o sinal de saída do LDR utilizamos a ferramenta de aquisição (Arduino), direcionando o sinal elétrico do sensor para uma de suas entradas analógicas. O Arduino é uma plataforma de hardware e software de código aberto projetada para aquisição e controle de sensores, motores e sistemas automatizados.

Com o uso dessa montagem é possível medir o período da passagem das hélices e assim calcular a frequência de rotação do motor concomitante a medida do $V_{rms}$ da f.e.m. induzida no multímetro. Na figura 3 temos 2 exemplos da aquisição da leitura da tensão de saída do LDR versus tempo da aquisição, para correntes de alimentação do motor de 110 e 630 mA. A tensão de saída do LDR é medida via porta analógica do Arduino. Para medida do tempo utilizamos no código a função *millis*, que retorna há quantos milissegundos o Arduino está ligado no momento da leitura da porta analógica.

O período $T_n$ é obtido pela diferença de tempo entre o número de sombras que a hélice da figura 2b faz no sensor LDR. No gráfico da figura 3 estas sombras são caracterizadas pelos mínimos de intensidade no eixo y. Portanto quando há a 5ª sombra, ou seja, a 1ª hélice volta a bloquear a luz novamente no LDR, obtemos o tempo decorrido equivalente ao período de revolução do motor, como destacado na figura 3 para a corrente de 110 mA.

A partir da medida do período a frequência linear e frequência angular são calculadas, respectivamente por:

$$f_n = 1/T_n \qquad (1)$$

$$\omega_n = 2\pi f_n \qquad (2)$$

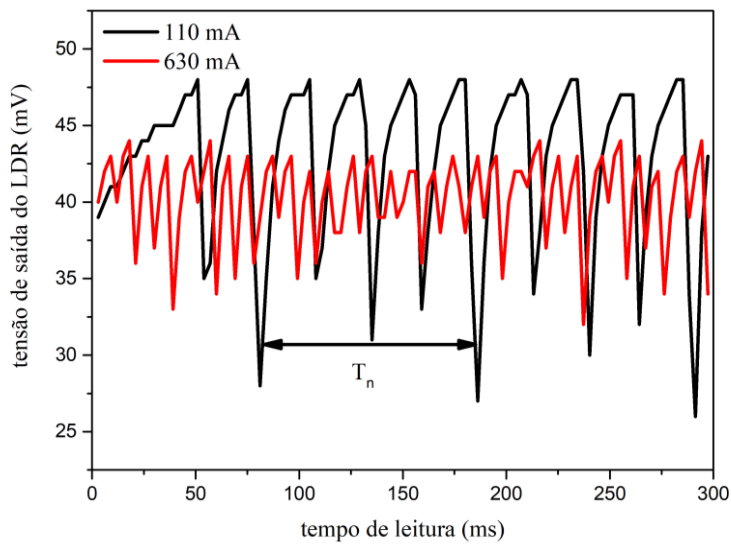

Figura 3: Aquisição dos valores do LDR em função do tempo de aquisição de leitura do Arduíno. Fonte: os autores.

Os valores obtidos para corrente de alimentação e frequência de rotação do motor e da f.e.m induzida no captador, encontram-se na Tabela 2. A f.e.m induzida é medida com um multímetro e por ser uma tensão alternada devido a variação periódica do fluxo do campo magnético, precisa ser medida no modo AC do multímetro. Esta medida fornece o valor da tensão elétrica em valor eficaz (*Root Mean Square Voltage*) e representa a média eficaz em um ciclo da f.e.m. induzida no captador. Para as medidas das incertezas, assumimos o valor fornecido pelo fabricante do multímetro para a corrente e o valor médio de variação da f.e.m (gerado principalmente pela trepidação da base de MDF). No caso das frequências de rotação, o procedimento adotado foi o de medir as diferenças entre os mínimos de intensidade da saída de leitura do LDR (ilustrados na figura 3 para correntes de 110 mA e 630 mA) e calcular a média e o desvio padrão da média. Para a incerteza relativa assumimos o valor da razão entre o desvio padrão/média.

Tabela 2 – Valores medidos para corrente de alimentação do motor DC, frequência de rotação do motor e f.e.m. Para a incerteza da medida da corrente, assumimos o valor fornecido pelo fabricante do multímetro. Para a f.e.m. usamos a variação média (aproximada) observada nas leituras do multímetro. Fonte: os autores.

| Corrente ($mA$) | Frequência ($Hz$) | f.e.m. ($mV$) |
|---|---|---|
| 110 $\pm$ 1% | 9,5 $\pm$ 4% | 654 $\pm$ 5% |
| 170 | 12,8 | 920 |
| 220 | 15,1 | 1105 |
| 290 | 17,1 | 1252 |
| 360 | 19,0 | 1371 |
| 430 | 20,8 | 1466 |
| 510 | 22,2 | 1565 |
| 570 | 23,8 | 1650 |
| 630 $\pm$ 1% | 25,6 $\pm$ 7% | 1719 $\pm$ 5% |

3. Análise
   3.1 A Lei de Faraday – Neumann - Lenz.

No início de 1831, Faraday desenvolveu estudos sobre a relação entre corrente elétrica e magnetismo, obtendo sucesso em produzir, pela primeira vez, o fenômeno de indução eletromagnética. Inicialmente Faraday vinha se dedicando ao estudo de figuras acústicas em sólidos e líquidos. No entanto, em agosto de 1831, Faraday descreveu pela primeira vez um anel feito de ferro doce em suas anotações (mais detalhes sobre a evolução do trabalho de Faraday podem ser encontrados no capítulo 2 da dissertação de mestrado de Dyego Soares de Lima (LIMA, 2020).

Faraday produziu um anel de ferro circular onde espiras de cobre foram enroladas em uma metade do anel onde 25 espiras foram separadas e isoladas por barbante e algodão. Elas foram produzidas por três extensões de fio que poderiam ser conectadas de forma única ou usadas de forma individual e esse lado do conjunto foi chamado de conjunto de espiras A. Isolado de A, procedimento semelhante foi feito no outro lado do anel e chamado de espiras B. As espiras do lado A foram conectadas a uma bateria. Ao fazer isso, Faraday detectou uma corrente elétrica no lado B. Ele encontrou o efeito que procurava, porém, percebeu que só ocorria no instante que conectava ou desconectava aos terminais da bateria. Esse experimento passou a ser conhecido como "*transformador de Faraday*" (LABAS 2016).

A partir deste ponto, ele conduziu outros ensaios para tentar reproduzir o fenômeno de indução magnética, em situações diferentes até que, ao final do ano de 1831, Faraday observou que utilizando um ímã em barra e movimentando-o pelo interior da bobina era possível gerar uma pequena corrente elétrica cujo sentido variava de acordo com o movimento produzido quando o ímã era colocado ou retirado. Faraday explicou o fenômeno usando um conceito chamado de linhas de força, mas a definição feita por ele não foi bem aceita por seus pares na época, por não ter uma base matemática (DIAS, 2004; DIAS e MARTINS, 2004).

Faraday também criou um dispositivo conhecido como disco de Faraday, que é considerado o primeiro gerador elétrico. Ele girou um disco condutor em um campo magnético, observando que isso gerava uma corrente contínua. Este experimento demonstrou como o movimento mecânico pode ser convertido em eletricidade. Os trabalhos originais de Michael Faraday sobre indução eletromagnética são encontrados em suas publicações e experimentos realizados entre 1821 e 1831. A descoberta e descrição da indução eletromagnética foram detalhadas por ele em várias notas e comunicações, particularmente em sua famosa série de experimentos de 1831. Os principais documentos que contêm essas contribuições são os "*Experimental Researches in Electricity*" publicados na revista *Philosophical Transactions of the Royal Society* em 1832 (FARADAY, 1832).

Em 1845, Franz Ernst Neumann fez a primeira formulação matemática da lei de indução, baseada nas análises de Faraday, e atualmente descrita como na equação da Lei de Indução de Faraday para a geração de uma força eletromotriz induzida (DIAS e MARTINS, 2004).

$$\mathcal{E} = -\frac{d\phi_B}{dt} \qquad (3).$$

A equação 3 demonstra que a força eletromotriz induzida $\mathcal{E}$ (fem) é proporcional a taxa de variação temporal do fluxo do campo magnético $d\phi_B$ (antes, linhas de força magnética). Posteriormente Heinrich Lenz também contribuiu com a formulação, atribuindo o sinal de negativo para que seja possível analisar o sentido da corrente induzida.

### 3.2 Relacionando as grandezas obtidas no experimento com o ensino da Lei de Faraday.

De uma forma geral os livros de física começam com uma abordagem histórica simplificada seguida de uma discussão qualitativa da lei de Faraday através de exemplos envolvendo, por exemplo, a produção de corrente elétrica a partir do movimento relativo entre uma espira e um imã permanente (HALLIDAY E RESNICK, 2016). À medida que a construção formal da lei de Faraday se concretiza, os livros convergem para os exemplos idealizados, nos quais a f.e.m induzida pode ser obtida diretamente a partir da equação 3.

Para exemplos cujas variações do fluxo do campo magnético sejam harmônicas destacamos duas situações, ilustradas na figura 4:

(i) Uma espira de área *A* que gira em torno de seu eixo (frequência angular $\omega$), em uma região de campo magnético uniforme $B_0$.

(ii) uma espira estática (parada) situada em uma região do espaço onde o campo magnético varia na forma harmônica $B_0 \cos(\omega t + \delta)$, onde $B_0$ é um campo uniforme e $\delta$ uma fase que não dependente do tempo.

[Digite aqui]

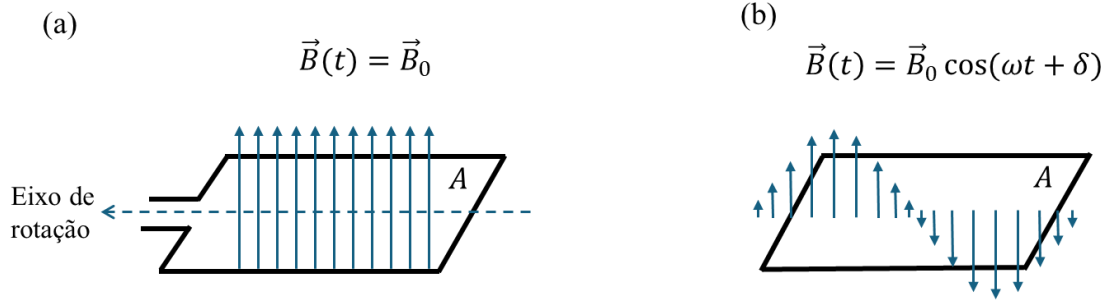

Figura 4: Exemplos de variações harmônicas do fluxo do campo magnético encontradas frequentemente em livros didáticos. (a) espira girando em um campo magnético uniforme, (b) espira estática em uma região de campo magnético dependente do tempo. Fonte: os autores.

Em ambos os casos, o fluxo do campo magnético é dado por:

$$\phi_B = \vec{B}.(\hat{n}A) \qquad (4).$$

Sendo $\hat{n}$ o vetor normal a área $A$. Desta forma, a f.e.m induzida $\mathcal{E}$ é dada segundo a equação 1 por:

$$\mathcal{E} = -\frac{d}{dt}(AB_0)\cos(2\pi f t + \delta)$$

$$= (2\pi AB_0)f\sin(2\pi f t + \delta) \qquad (5).$$

Para estabelecer uma relação entre os resultados do experimento listados na tabela 2 e a abordagem teórica da equação 5 devemos analisar o comportamento da f.e.m induzida com o aumento da frequência, ilustrada na figura 5 para valores de frequências entre 9,5 e 26,6 Hz. Lembrando que a frequência de rotação do motor está ligada à taxa de variação do fluxo do campo magnético na face do captador.

[Digite aqui]

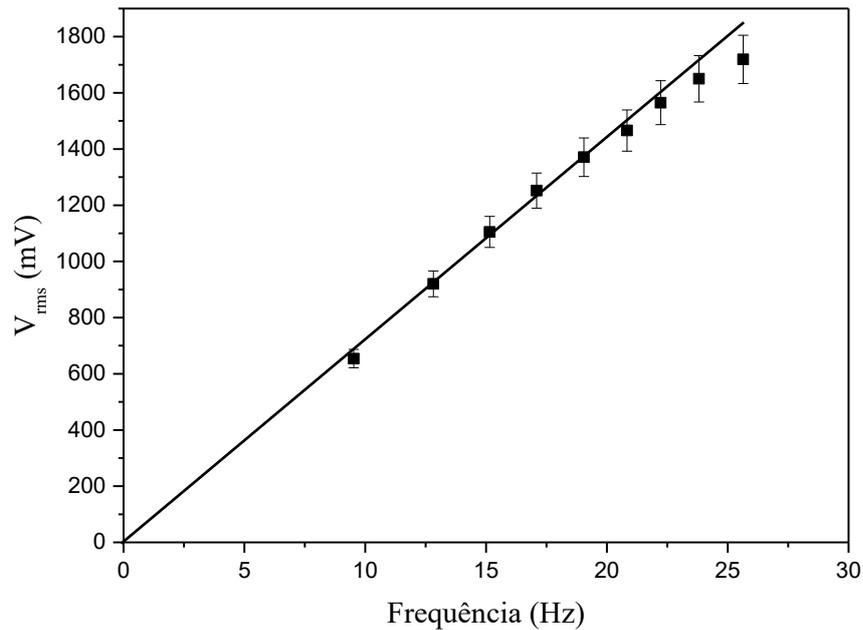

Figura 5: medidas obtidas experimentalmente com o kit da f.e.m. induzida no captador vs. Frequência de rotação do motor. A linha cheia indica um ajuste linear dos dados obtidos. Fonte: os autores.

Interessante observar que entre 10 e 20-25 Hz a relação entre a f.e.m. induzida medida e a frequência de rotação é linear. Neste momento podemos comparar o valor medido $V_{rms}$ do experimento e o resultado de $\mathcal{E}$ para variações harmônicas da equação 5.

Esta correlação deve ser feita comparando o valor medido $V_{rms}$ com o valor médio quadrático de $\mathcal{E}$ definido por $\mathcal{E}_{rms} = \sqrt{\langle \mathcal{E}^2 \rangle}$, onde:

$$\langle \mathcal{E}^2 \rangle = \frac{1}{T} \int_0^T \mathcal{E}(t)^2 \, dt \qquad (6)$$

Realizando a média quadrática no período $T$, para o resultado da equação 5, obtemos $\mathcal{E}_{rms}$:

$$\mathcal{E}_{rms} = \left(\frac{2\pi A B_0}{\sqrt{2}}\right) f \qquad (7)$$

Este resultado é muito importante porque demonstra que teoricamente, em uma situação de variação harmônica do fluxo do campo magnético, $\mathcal{E}_{rms}$ varia linearmente com a frequência. Experimentalmente, o resultado de $V_{rms}$ ilustrado na figura 5, também apresenta um comportamento linear com a frequência entre 10 e 20 - 25 Hz.

Por se tratar de um experimento e não um exemplo idealizado, há diversas considerações que devem ser discutidas com os estudantes acerca da expectativa de termos concordância entre experimento e modelos teórico. Em particular a limitação do captador da guitarra em separar a variação dos campos magnéticos de diversos imãs à medida que a frequência de rotação aumenta.

4   Comentários finais.

A busca por engajamento dos alunos em sala de aula nas disciplinas de Física tem sido um tema constante de estudos e iniciativas, voltados a propor abordagens que integrem diferentes referenciais teóricos. Como parte dessa proposta, foi desenvolvido um *kit* para ser levado à sala de aula que utiliza o captador da guitarra como sensor da variação do fluxo de campo magnético, permitindo explorar a Lei da Indução de Faraday de maneira prática e adaptável a diferentes metodologias educacionais. Nesse contexto, o *kit* proposto se destaca pela:

i) Montagem portátil que pode ser levado a sala de aula convencional sem a necessidade de estar em um laboratório didático. Custo acessível dos itens utilizados não havendo necessidade de utilizar um osciloscópio.
ii) O modelo *single coil* do captador da guitarra se mostrou um bom sensor para medir a variação do fluxo do campo magnético. Em particular pelo fato da f.e.m. induzida ser suficientemente grande para ser medida utilizando um multímetro, sem necessidade de amplificação.
iii) Aprendizado ativo: Os alunos têm a oportunidade de coletar dados e analisar as grandezas físicas relacionadas aos exemplos apresentados em livros didáticos.
iv) Contextualização. A importância da lei de indução, a partir do entendimento de como funciona o primeiro passo (elétrico) na tradução da vibração mecânica em som. A partir daí o professor pode explorar a interdisciplinaridade da física com outras áreas, como a música e a engenharia. Eles podem investigar posteriormente como o projeto dos captadores influencia o timbre e o som da guitarra, bem como o papel da eletricidade e do magnetismo na amplificação do som.